\documentclass[prb,showpacs,twocolumn,preprintnumbers,amsmath,amssymb,floatfix]{revtex4}
\usepackage[dvips]{graphicx}
\usepackage[latin1]{inputenc}
\begin{document}
\draft

\newcommand{\pb} {Pb$_2$VO(PO$_4$)$_2$\,}
\newcommand{\bac} {BaCdVO(PO$_4$)$_2$\,}
\newcommand{\lisi} {Li$_2$VOSiO$_4$\,}
\newcommand{\etal} {{\it et al.} }
\newcommand{\ie} {{\it i.e.} }
\newcommand{\aucr}{CeCu$_{5.9}$Au$_{0.1}$ }
\newcommand{\auaf}{CeCu$_{5.2}$Au$_{0.8}$ }
\newcommand{\aux}{CeCu$_{6-x}$Au$_{x}$ }
\newcommand{\ip}{${\cal A}^2$ }

\hyphenation{a-long}

\title{Fluctuations and correlations in a frustrated $S=1/2$ square lattice with
competing ferromagnetic and antiferromagnetic interactions: a $\mu$SR study}

\author{P. Carretta$^1$, M. Filibian$^1$, R. Nath$^{2,*}$, C. Geibel$^2$, P.J.C. King$^3$}
\affiliation{$^1$ Dipartimento di Fisica ``A.Volta", University of Pavia - CNISM,  I-27100 Pavia, Italy}
\affiliation{$^2$ Max Planck Institute for Chemical Physics of Solids, 01187 Dresden, Germany} \affiliation{$^3$
ISIS Facility, Rutherford Appleton Laboratory, Chilton, Didcot, OX11 0QX, UK} \altaffiliation[Present address:
]{Ames Laboratory and Department of Physics and Astronomy, Iowa State University, Ames, Iowa 50011, USA}

\widetext

\begin{abstract}

Zero and longitudinal field $\mu$SR measurements in \pb and \bac, two prototypes of the frustrated $S=1/2$ square
lattice model with competing ferromagnetic and antiferromagnetic interactions, are presented. Both systems are
observed to undergo a phase transition to a long-range magnetic order at $T_N\simeq 3.46$ K, for \pb, and at
$T_N\simeq 0.99$ K, for \bac. In \pb both the temperature dependence of the order parameter and the longitudinal
relaxation rate above $T_N$ are consistent with a two-dimensional XY model. On the other hand, for \bac, which
lies very close to the magnetically disordered region of the phase diagram where a bond-nematic order was
predicted, a peculiar logarithmic increase of the relaxation is observed above $T_N$. In both systems a rather
broad distribution of internal fields at the muon sites is noticed below $T_N$. The origin of this distribution is
discussed in the light of the $\mu$SR experiments already performed on $S=1/2$ frustrated antiferromagnets on a
square lattice.

\end{abstract}

 \maketitle
\narrowtext

\section{Introduction}

Strongly correlated electron systems with competing interactions are known to show rather rich phase diagrams,
with crossovers or phase transitions which depend on the relative magnitude of the competing energy scales
\cite{HF}. In certain cases rather complex scenarios are observed together with the insurgence of novel phases of
matter, a situation typically found in frustrated magnets where, for instance, spin ice or spin liquid
ground-states arise.\cite{Diep} In these insulating systems either the geometry of the underlying spin lattice or
the geometry of the interactions causes the suppression of the long-range magnetic order and the onset of exotic
phases. For instance, for a frustrated $S=1/2$ square lattice with competing ferromagnetic and antiferromagnetic
interactions \cite{Shan}(hereafter QFFMSL for short) it has been proposed that for exchange couplings yielding to
the disappearance of long-range magnetic order, a nematic order could be present.\cite{Nema} QFFMSL are
characterized by a ferromagnetic nearest neighbour (n.n.) exchange coupling $J_1$, along the sides of the square
spin lattice, competing with the antiferromagnetic next nearest neighbour (n.n.n.) coupling $J_2$, along the
diagonals. For a frustration ratio $-0.4\geq J_2/J_1\geq -0.7$ long-range magnetic order should disappear and a
nematic order of spin bonds should arise.\cite{Nema} In this phase, the individual spins do not carry any
orientational order but the traceless tensor
\begin{equation}
\mathcal{O}^{\alpha\gamma}= \frac{1}{2}(S_i^{\alpha}S_j^{\gamma}+S_i^{\gamma}S_j^{\alpha}) -
\frac{1}{3}\delta^{\alpha\gamma}\langle\vec S_i \vec S_j\rangle\,\,\,\, ,
\end{equation}
with $\alpha,\gamma= x, y, z$ and $i, j$ running over spin lattice sites, does order and gives rise to stripy like
correlations in the plane. This interesting theoretical result lacks of any experimental confirmation, basically
because there is no material characterized by a frustration ratio $J_2/J_1\simeq -0.5$. Nevertheless, in the last
years there has been a considerable effort to synthesize novel vanadates, containing V$^{4+}$ $S=1/2$ ions, which
have exchange couplings in this region of the $J_1-J_2$ phase diagram. Several prototypes QFFMSL have been
synthesized \cite{Geibel}, with ratios $J_2/J_1$ ranging from $-1.8$ for \pb to -0.9 for \bac \cite{Nath}, very
close to the boundary between the long-range magnetic order and the nematic order. On the other hand, it has been
recently pointed out  by Tsirlin and Rosner \cite{Rosn} that in these compounds the n.n. and n.n.n. exchange
couplings along different sides and diagonals of the square lattice, respectively, might be inequivalent due to
the buckling and stretching of the magnetic layers. The difference between the exchange couplings appears to be
less pronounced for \bac than for \pb.

\begin{figure}[h!]
\vspace{6cm} \includegraphics{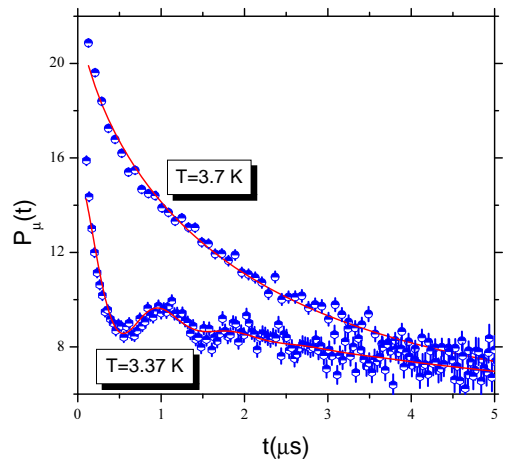}  \caption{Time evolution of the
ZF$\mu$SR asymmetry in \pb at two representative temperatures, above and below $T_N$. The solid lines are best
fits according to Eq. 2 (data at $T= 3.7$ K) and 3 (data at $T= 3.37$ K) in the text.}\label{Fig1}
\end{figure}

In order to study the evolution of the local microscopic properties of QFFMSL upon decreasing the ratio $J_2/J_1$
towards the critical value $J_2/J_1\simeq -0.7$ and how they compare to the ones of frustrated $S=1/2$ square
lattice systems with competing antiferromagnetic interactions \cite{Car1,Rosn2} (QFAFSL, i.e. for $J_2/J_1> 0$),
we have performed zero field (ZF) and longitudinal field (LF) $\mu$SR experiments on \bac and \pb powders. It was
found that although both systems are characterized by a magnetically ordered ground-state below $T_N$, significant
differences with respect to the behaviour observed for QFAFSL are present. The correlated spin dynamics in \pb for
$T> T_N$ and the temperature dependence of the magnetic order parameter are both consistent with a two-dimensional
(2D) XY model. On the other hand, in \bac a rather peculiar behaviour is observed, with a logarithmic increase of
the longitudinal muon relaxation rate ($\lambda$) on cooling for $T> T_N$. This behaviour is reminiscent of the
one observed in one-dimensional systems and might suggest the onset of bond-nematic correlations above $T_N$.

\section{Technical aspects and experimental results}
A polycrystalline sample of \pb was synthesized by solid state reaction technique using PbO ($99.99$\%),
(NH$_{4}$)H$_{2}$PO$_{4}$ ($99.9$\%), and VO$_{2}$ ($99.99$\%) as starting materials. In the first step, the
intermediate compound Pb$_{2}$P$_{2}$O$_{7}$ was prepared firing the stoichiometric mixtures of PbO and
(NH$_{4}$)H$_{2}$PO$_{4}$ in air at $750$ $^{\circ}$C for $24$ hours. In the second step, the intermediate product
was mixed with VO$_{2}$ in appropriate molar ratio and heated for $48$ hours at $680$ $^{\circ}$C in dynamic
vacuum with one intermediate grinding and pelletization. Synthesis of \bac was done following the same procedure
reported in Ref.~\onlinecite{Nath}. Single phase materials were confirmed by x-ray diffraction performed with a
STOE powder diffractometer (CuK$_{\alpha}$ radiation). Both materials, \pb and \bac, contained a minor (2-3\%)
fraction of unreacted diamagnetic phosphates Pb$_2$P$_2$O$_7$ and BaCdP$_2$O$_7$, respectively. These impurities
are non-magnetic and, therefore, are irrelevant for the discussion to be presented in the following.

ZF and LF$\mu$SR measurements were performed on EMU and MUSR beam lines at ISIS pulsed muon facility, using $29$
MeV/c spin-polarized muons. The powders of \pb and \bac were pressed and attached to a silver sample-holder, whose
background contribution ($B$) to the muon asymmetry was determined from the slowly decaying part of the
longitudinal polarization. We checked that there were no history dependent effects associated with a poor sample
thermalization onto the dilution fridge cold plate.

\begin{figure}[h!]
\vspace{6cm} \includegraphics{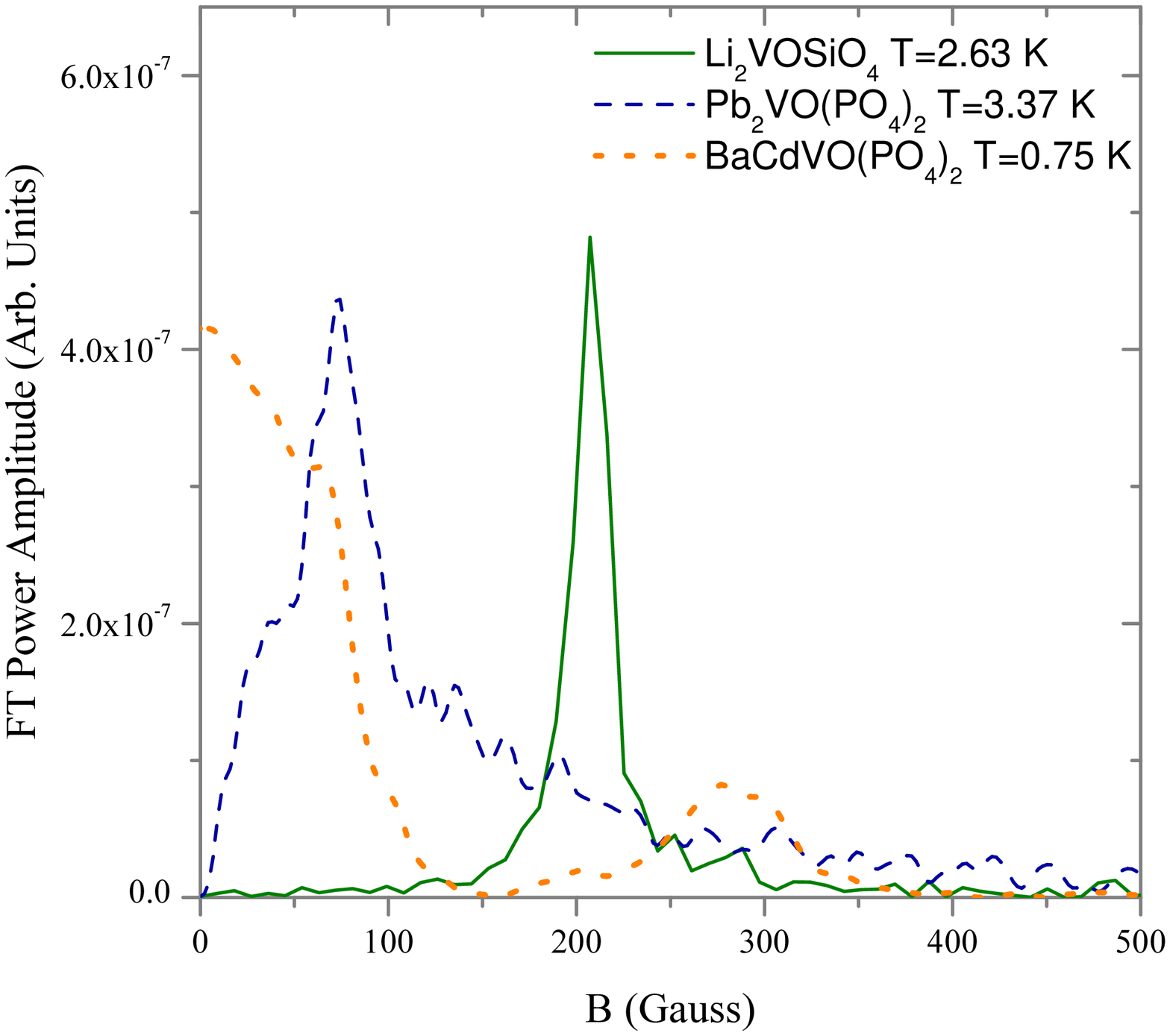}  \caption{Power of the Fourier
transform of the oscillating component of the ZF$\mu$SR asymmetry for \pb , \bac and \lisi, for $T< T_N$. It is
evident that while in \lisi \cite{Carmu} a well defined internal field is present, in \pb and \bac a broad
distribution of internal fields is probed by the muons. }\label{Fig2}
\end{figure}

In \pb at temperatures above $T_N= 3.46\pm 0.01$ K the decay of the muon polarization $P_{\mu}(t)$, either in zero
(Fig. 1) or in a longitudinal field,  could be fitted by the expression
\begin{equation}
P_{\mu}(t)= A exp(-(\lambda t)^{\beta}) + B\,\,\, .
\end{equation}
In ZF, for $T\gg T_N$, $\beta$ was found to increase slightly above unity which indicates that at high temperature
a non-negligible contribution to the relaxation arises from the field distribution generated by the dipolar
interaction with the nuclei, which typically gives a Gaussian decay.\cite{Schenck} On the other hand, when a LF of
1 kGauss was applied in order to quench the nuclear dipole contribution, $\beta$ was found to decrease from unity
to about $0.7$ on approaching $T_N$.

\begin{figure}[h!]
\vspace{8.5cm} \includegraphics{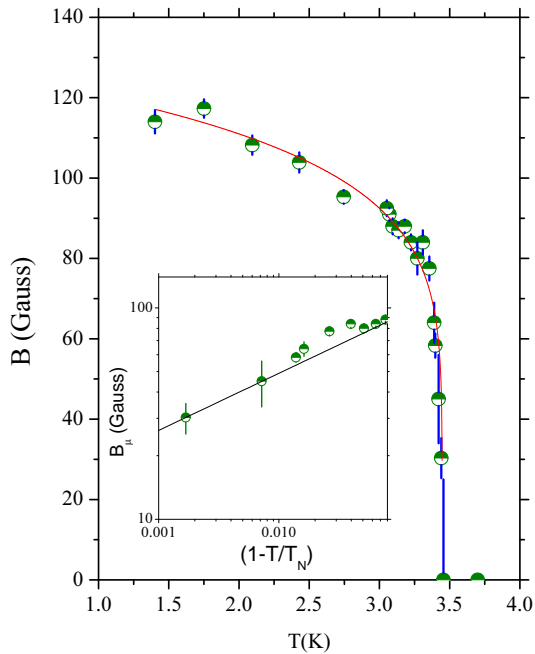} \caption{Temperature dependence
of the internal field at the muon in \pb as derived from ZF$\mu$SR experiments. In the inset the same data are
reported as a function of the reduced temperature for $T\rightarrow T_N$. The solid line is the critical behaviour
expected for a 2D XY model. }\label{Fig3}
\end{figure}

Below $T_N$ ZF experiments showed the insurgence of oscillations in the muon asymmetry (Fig. 1) which clearly
indicate the presence of a long-range magnetic order, which generates an internal field at the muon site
$B_{\mu}$. Accordingly one observes a precessional signal at a frequency $\omega_{\mu}=\gamma_{\mu}B_{\mu}$, with
$\gamma_{\mu}$ the muon gyromagnetic ratio. Nevertheless, the initial decay of the muon polarization can hardly be
fitted by assuming a well defined field at the muon. On the other hand, one can fit $P_{\mu}(t)$ over a broad
temperature range below $T_N$ with the expression
\begin{equation}
P_{\mu}(t)= A_1 exp(-\sigma t)J_0(\gamma_{\mu}B_{\mu}t) + A_2 exp(-(\lambda t)^{\beta}) + B\,\,\, ,
\end{equation}
where $J_0(x)$ is the zero-th order Bessel function of the first kind, which characterizes the decay of the muon
polarization in the presence of a distribution of internal fields \cite{SDW} which is further broadened by the
decay constant $\sigma$. In fact, if one performs the Fourier transform of the oscillating component of
$P_{\mu}(t)$ one finds quite a broad spectrum (Fig. 2). Upon decreasing the temperature below $T_N$ one observes a
loss of the initial asymmetry, possibly due to a further broadening of the field distribution. The temperature
dependence of $B_{\mu}$, derived by fitting the ZF decay of the muon asymmetry with Eq. 3 is reported in Fig. 3.
In Fig. 4 the temperature dependence of the longitudinal relaxation rate $\lambda$ is reported. A clear divergence
of the relaxation rate is observed at $T_N$.

\begin{figure}[h!]
\vspace{8.5cm} \includegraphics{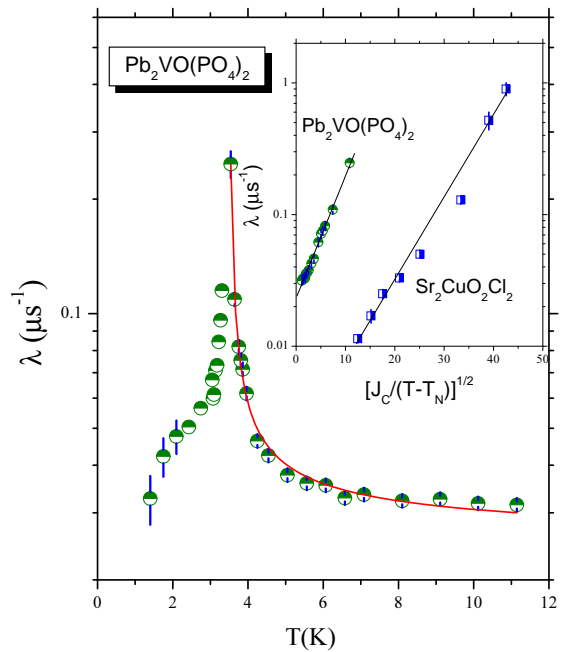} \caption{Temperature dependence
of the muon longitudinal relaxation rate $\lambda$ in \pb . Above $T_N$ $\lambda$ was derived in a field of 1000
Gauss, while below $T_N$ it was estimated from the non-oscillating component of the muon asymmetry. The solid line
shows the behaviour expected for a 2D XY model. In the inset the same data are reported for $T\rightarrow T_N$ as
a function of $\sqrt{J_c/(T-T_N)}$, together with the data derived for Sr$_2$CuO$_2$Cl$_2$, for a LF $H= 800$
Gauss along the $c$ axes. $J_c= 10.4$ K for \pb and $J_c= 1450$ K for
Sr$_2$CuO$_2$Cl$_2$.\cite{Borsa}}\label{Fig4}
\end{figure}

Also for \bac at temperatures above $T_N= 0.99\pm 0.01$ K the decay of the muon polarization can be nicely fitted
with Eq. 2, with an exponent  decreasing from $\beta= 1$ for $T\simeq 10$ K to $\beta= 0.6$ on approaching $T_N$
(Fig. 5). On the other hand,  although in ZF below $T_N$ oscillations in $P_{\mu}(t)$ are observed, confirming the
onset of a long-range magnetic order \cite{Nath}, the decay of the muon polarization can hardly be fitted by Eq. 3
and a much broader distribution of internal fields is present. In fact, by looking at the Fourier transform of the
oscillating component of the muon asymmetry (Fig. 2) one clearly notices that a significant distribution of
internal fields is present, much broader than in QFAFSL \cite{Carmu}. Still, it is possible to fit the slowly
decaying non-oscillating component of $P_{\mu}(t)$ with a stretched exponential in order to derive the temperature
dependence of $\lambda$ over all the explored temperature range. Again a peak is present at $T_N$ (Fig. 6).
However it is noticed that for $T> T_N$ the increase of $\lambda$ on cooling is rather different from the one
observed in \pb .

\begin{figure}[h!]
\vspace{6cm} \includegraphics{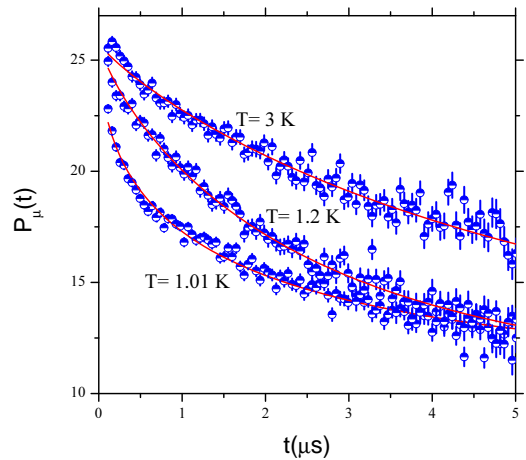} \caption{Time evolution of the
ZF$\mu$SR asymmetry in \bac at three representative temperatures above $T_N$. The solid lines are best fits
according to Eq. 2.}\label{Fig5}
\end{figure}

\section{Discussion}

First we shall address the temperature dependence of the muon longitudinal relaxation rate and the temperature
dependence of $B_{\mu}$ in \pb . The temperature dependence of $B_{\mu}$ corresponds to the one of the order
parameter. In fact, in zero-field the local field at the muon can be written as $\vec B_{\mu}= \mathcal{A} <\vec S
>$ \cite{Schenck}, where $\mathcal{A}$ is the hyperfine coupling tensor describing the coupling of the muon with
the surrounding V$^{4+}$ $S=1/2$ spins and $<\vec S>$ the corresponding spontaneous spin polarization. In spite of
the distribution of local fields evidenced in Fig. 2, which yields some uncertainty in the estimate of $B_{\mu}$,
the basic trend of $B_{\mu}(T)$ is reminiscent of the one observed for \lisi QFAFSL.\cite{Carmu} In this latter
compound it has been observed that for $T\rightarrow T_N$ the order parameter increases according to the critical
power law $B_{\mu}\propto (1- T/T_N)^{\beta_c}$, with a critical exponent $\beta_c\simeq 0.235$, the one predicted
for a 2D XY system.\cite{Bram} If one reports $B_{\mu}$ data for \pb in the proximity of $T_N$ as a function of
the reduced temperature $(1- T/T_N)$ (see the inset of Fig. 3), one realizes that also for \pb $\beta_c\simeq
0.235$.
\begin{figure}[h!]
\vspace{8.5cm} \includegraphics{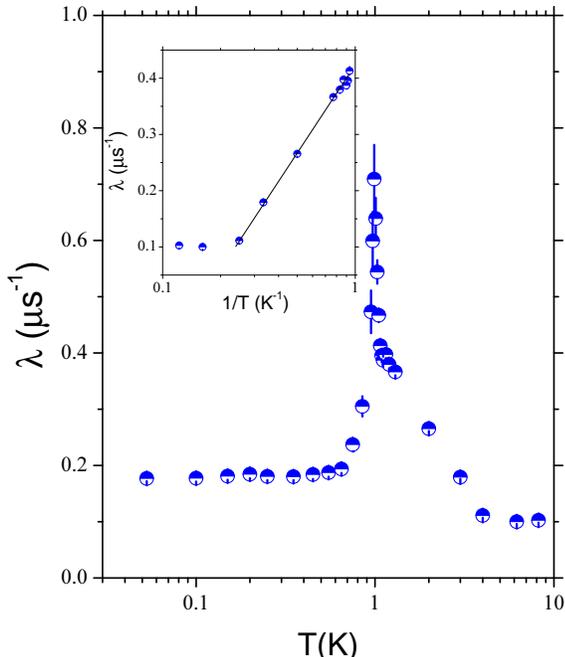} \caption{Temperature dependence
of the zero-field muon longitudinal relaxation rate $\lambda$ in \bac . In the inset the same data are reported
vs. $1/T$ in a linear-log scale in order to evidence the logarithmic increase of $\lambda$ above $T_N$.
}\label{Fig6}
\end{figure}

The temperature dependence of the muon longitudinal relaxation rate above $T_N$ originates from the one of the
in-plane correlation length. In fact, if we neglect the nuclear contribution to the relaxation which is
temperature independent and in anyway, quenched by the application of a longitudinal field, the critical increase
of $\lambda$ on approaching $T_N$ is of dynamical nature and associated with spin-lattice relaxation processes.\cite{Schenck} Then one can write
\begin{equation}
\lambda= \frac{\gamma_{\mu}^2}{2N}\sum_{\vec q,\alpha}|A_{\vec q}|^2 S^{\alpha\alpha}(\vec q ,\omega_{\mu})\,\,\,
,
\end{equation}
where $S^{\alpha\alpha}(\vec q ,\omega_{\mu})$ are the components of the dynamical structure factor at the muon
Larmor frequency and $A_{\vec q}$ is the form factor describing the hyperfine coupling of the spin excitations at
wave-vector $\vec q$ with the nuclei. By resorting to scaling arguments, which are expected to be valid for
$T\rightarrow T_N$, one finds that for a 2D spin system $\lambda\propto \xi^z$, with $\xi$ the in-plane
correlation length and $z$ the dynamical scaling exponent.\cite{Carcsi} Considering that \pb lyes in the sector of
the $J_1-J_2$ phase diagram where the antiferromagnetic $J_2$ coupling is larger, one could at first take for
$\xi$ the temperature dependence expected for a $S=1/2$ 2D Heisenberg antiferromagnet on a square lattice. Namely,
$\xi\propto exp(2\pi\rho_s/T)$ \cite{CHN}, with $\rho_s$ the spin-stiffness constant, which is possibly reduced
with respect to the value $\rho_s= 1.15 J/2\pi$ expected for a non-frustrated systems. Accordingly one should
observe that $\lambda\propto exp(2z\pi\rho_s/T)$. However, we have found that this expression can fit the data
only over a limited temperature range, no matter which value one takes for $\rho_s$. On the other hand, the
temperature dependence of the order parameter below $T_N$ suggests considering for $\xi$ the form predicted for a
2D XY model \cite{Ding}, namely $\xi\propto exp[B_{XY}/(T- T_N)]^{1/2}$, where $B_{XY}$ is of the order of
magnitude of the exchange coupling constant. Then, one should find
\begin{equation}
\lambda= c \times e^{[z^2 B_{XY}/(T- T_N)]^{1/2}}\,\,\, .
\end{equation}
Indeed, \pb data can be nicely fitted with Eq. 5 over almost all the explored temperature range (Fig. 4). It is
worth comparing the behaviour observed for \pb with the one found for a prototype of the $S=1/2$ 2D XY model,
Sr$_2$CuO$_2$Cl$_2$.\cite{Borsa} In Fig. 4 we compare the temperature dependence of $\lambda$ for \pb and for
Sr$_2$CuO$_2$Cl$_2$, reported as a function of $\sqrt{J_c/(T- T_N)}$, with $J_c= \sqrt{J_1^2 + J_2^2}$ an
effective exchange coupling. One notices that for both systems Eq. 5 is followed very well.

Now we shall address the temperature dependence of the order parameter describing the \bac ground-state. A broad
distribution of local fields at the muon site, even larger than the one found for \pb, is noticed (Fig. 4). This
field distribution should not be associated with sample inhomogeneities since the phase transitions detected by
$\lambda$ or specific heat measurements \cite{Nath} appear rather sharp. On the other hand, it should be mentioned
that in QFAFSL as \lisi, where the buckling along the $a$-axis is not present and the unit cell is tetragonal,
such a broad distribution is absent \cite{Carmu,Bombardi}(Fig. 4). Thus, at first sight one could think that the
different structure could yield a distribution of muon sites, leading to a distribution of hyperfine coupling
constants which could justify the broadening of the Fourier transform spectra. However, it is rather difficult to
associate such a broad distribution with the presence of a discrete number of $\mu^+$ sites and moreover the
distribution is larger in \bac where the $ab$ plane is less distorted.\cite{Rosn} Furthermore, even if a priori
the presence of several inequivalent muon sites cannot be excluded, we mention that in oxides with superstructures
and much larger unit cells, as Bi$_2$Sr$_2$YCu$_2$O$_{8-\delta}$ \cite{Bimu} and Sr$_{14}$Cu$_{24}$O$_{41}$
\cite{Sr14} for instance, just up to two muon sites are observed.

If the significant broadening in the Fourier transform is not due to a distribution of inequivalent probes it
means that $<\vec S>$ varies from lattice cell to lattice cell, namely either the underlying magnetic lattice is
not commensurate with the crystal lattice structure or domains are present at the microscopic level. In this
respect it is instructive to compare our results to neutron scattering ones.\cite{NS} Although neutron scattering
experiments in \pb indicate the presence of a columnar order, a broad background distribution of the energy
integrated dynamical structure factor is noticed in $q$-space, even well below T$_N$. \cite{NS} The origin of this
broad distribution could suggest the presence of small size domain like structures. Indeed \pb and \bac are in a
sector of the $J_1-J_2$ phase diagram where a two-fold degenerate magnetic ground-state is present \cite{Chandra}
and one could speculate that under certain conditions the coexistence of domains of both phases, characterized by
different magnetic wave vectors, could be observed. Although, we cannot draw a clear conclusion on the origin of
the broadening of the spectra, it is clear that in \bac and \pb a larger degree of structural and possibly of
magnetic disorder is present with respect to QFAFSL as \lisi.

In \bac the application of a moderate magnetic field is observed to affect significantly the spectrum of the
excitations and the specific heat.\cite{Nath} Therefore, in order to probe the intrinsic low-energy excitations of
this system we have performed zero-field relaxation experiments. Under these conditions an additional contribution
to the decay of the muon polarization, associated with the dipolar interaction with the nuclei, might be present.
Nevertheless, the functional form of $P_{\mu}(t)$ suggests that in \bac this contribution is less significant than
in \pb, where the nuclear contribution to the relaxation is already minor for $T\rightarrow T_N$. In fact, in this
latter compound $\lambda$ is observed to diverge for $T\rightarrow T_N$ according to Eq.5 also in ZF.

On the other hand, the temperature dependence of the zero-field relaxation rate in \bac shows a rather peculiar
behaviour, not observed in \pb. At high temperature, for $T> 4$ K, $\lambda$ is temperature independent, as
expected for a non-correlated spin system. On decreasing the temperature towards $T_N$, one would expect for a 2D
system, on the basis of Eq. 4, an exponential growth of $\lambda$. Remarkably $\lambda$ is found to increase
logarithmically, i.e. $\lambda\propto ln(1/T)$ (Fig. 6), a trend which is rather characteristic of one-dimensional
systems.\cite{Taki} This behaviour should not be related to the filtering of the critical fluctuations by the form
factor in Eq. 4.\cite{Carcsi} In fact, if this was the case one should expect the cancellation of the magnetic
field at the muon site below $T_N$, at variance with the experimental findings (Fig. 2). On the other hand, since
\bac should lie very close to the boundary with the bond-nematic phase \cite{Nath,Nema}, one could expect that
nematic correlations of stripy character arise above $T_N$ and yield a logarithmic increase of $\lambda$ with
decreasing temperature. This occurs until the XY anisotropy and/or the interlayer coupling cause the crossover to
a three-dimensional long-range order at a finite temperature resulting in an abrupt increase of $\lambda$ very
close to $T_N$ (see Fig. 6).

\section{Conclusions}

In this manuscript we have presented new $\mu$SR measurements in two prototypes of QFFMSL: \pb and \bac. In both
compounds a broad distribution of local fields at the muon site is evidenced and is tentatively associated either
with an incommensurate magnetic order or with the formation of mesoscopic domains. In \pb the overall temperature
dependence of the longitudinal relaxation rate and of the order parameter below $T_N$ are both consistent with a
2D XY model. On the other hand, in \bac a rather peculiar behaviour of the longitudinal relaxation rate was
evidenced, which can be hardly associated with conventional 2D spin correlations. The logarithmic increase of
$\lambda$ on cooling suggests the onset of one-dimensional correlations which might be related to the insurgence
of novel type of correlations in this QFFMSL.

\section{Acknowledgements}
We are thankful to A.~A.~Tsirlin for his help during sample preparation and for the enlightening discussions.
L.L.Miller is gratefully acknowledged for providing the Sr$_2$CuO$_2$Cl$_2$ single crystal. Financial support from
Fondazione Cariplo 2008-2229 research funds is gratefully acknowledged.



\end{document}